\documentstyle[epsf,prl,aps,twocolumn]{revtex}
\begin{document}
\draft
\title{ Waves in Schwarzschild spacetimes: how strong can be
imprints of the spacetime curvature}
\author{Janusz Karkowski$^*$, Krzysztof Roszkowski$^*$,
 Zdobys\l aw \'Swierczy\'nski$^{+}$ and  Edward Malec$^{*,**}$}
\address{ $^*$ Institute of Physics,  Jagiellonian University,
30-059  Cracow, Reymonta 4, Poland.}
\address{$^{**}$ Max-Planck-Institut f\"ur Gravitationsphysik,
 Albert-Einstein-Institut, Am M\"uhlenberg 1, Golm, Germany.}
\address{$^{+}$ Pedagogical University, Cracow, Podchor\c a\.zych 1, Poland.}
\maketitle
\begin{abstract}
Emitted radiation can be reprocessed in curved spacetimes, due to the
breakdown of the Huyghens principle. A maximization procedure for the
energy diffusion allows one to obtain wave packets (gravitational and
electromagnetic) that are particularly strongly backscattered. Examples
are  shown with the backscattered part exceeding by one order
remnants of initial signals. A robust ringing can be observed,
with amplitudes exceeding leftovers of the main radiation pulse.
The analysis of the obtained results allows one to set
 demands on some parameters in  the numerical description of a
realistic process of the collapse of two black holes.
\end{abstract}

\section{Introduction}

It is known essentially since the times of Hadamard \cite{Hadamard} that
  curved spacetimes can affect the propagation of waves. The breakdown of
the Huyghens principle \cite{Hadamard} (or the backscatter, a name adopted by
the general relativity community after de Witt and Brehme \cite{DeWitt})
can influence both the energy and the energy flux of a wave signal. The
backscatter can leave its imprints on the frequency spectrum and can
affect the  transmission time. The manifestations of this
effect are the so-called tails and,  most impressively,
the quasinormal modes (QNM thereafter). The literature on the
backscattering and related phenomena is quite extensive  -- see
(\cite{lit} -- \cite{Stalker}) and numerous references therein.

The QNM's  have some features of the scattering-type solutions and they
have been studied in the context of general relativity for more than
three decades \cite{Vishveshvara}.
 Many of their characteristics
are well known for black holes \cite{Leaver}, for instance their  (complex)
frequency spectrum. An observer located at a
 fixed space position would find that QNM's
oscillate with amplitudes decreasing exponentially in time.
The oscillation periods and the damping exponents are the real and the imaginary
 parts of a frequency, respectively.  They depend only on a few global
characteristics of black holes -- their asymptotic
  mass, angular momentum and/or
global electric charge. Therefore their
 identification in an observed wave spectrum
would unambiguously identify a black hole (and in fact provide an argument,
closest to the direct observation, in favour of the existence of black holes).
   Extensive reviews are presented in \cite{Kokkotas}
and \cite{Nollert}. Tail terms have been studied
  in 1970's beginning from Price \cite{Price1970}
 but interest in them has  again  revived recently \cite{Stalker}.

The spectra of QNM's and the decay exponents of the tails are universal,
independent of initial data, but the very existence of QNM's
and their amplitudes (as well of the tails)   do depend on initial
wave conditions. The main aim of this paper is to show the {\bf strongest
imprints of the spacetime curvature  that are present in the form of QNM's in
a propagating wave}. The implementation of this task requires the
separation of the genuine geometric effects from those  being built into
initial data -- notice that even  in the Minkowski spacetime one can easily
form a QNM-like structure by producing suitable initial data.
The simplest possibility is to consider
the purely backscattered part of the initial radiation, which is absent
in the Minkowski geometry but which always exists in a curved  spacetime.

It would be meaningless to try to accomplish our aim  by the method of
"trial and error" --  by selecting  at random  various initial wave
configurations from the ocean of all possible data. Rather one should focus on
``extremal'' in some sense initial data that can generate, in the first
instance, ``extremal'' asymptotic templates, but also
can set bounds on some parameters used in the numerical descriptions. In the
present paper we follow the second strategy, using as a guiding principle the
idea of extremizing the so-called diffusion parameter \cite{Kark2002} and
addressing following issues. First, we estimate the maximal strength of the
backscatter. The corresponding profiles of initial wave packets  are found
to favour vigorous ringing and/or strong deformation of  initial signals.
Second, and in relation with the former point,
we obtain  information  on the process of  taking waveforms
from  specific properties of the backscattered radiation.
The order of the rest of this paper is as follows.
 Sec. 2 provides basic information
on the wave equations. Sec. 3 describes in detail the procedure of
maximizing the  diffusion parameter and
 shows exemplary initial data for the wave
evolution. Sec. 4 reviews  some representative examples of wave templates.
  In Sec.  5  we again review those features of the
numerical examples that could be useful for the numerical relativists
dealing with the full nonlinear description of the collapse of two black holes.
Sec. 6 summarizes main conclusions.

\section{Basic definitions and  concepts}

\subsection{Equations}

The spacetime geometry is defined by the line element
\begin{equation}
ds^2=-\eta_Rdt^2+{dR^2\over \eta_R}+R^2d\Omega^2,
\label{1}
\end{equation}
where $t$ is a time coordinate, $R$ is the radial areal coordinate,
$\eta_R=1-{2m\over R}$ and $d\Omega^2=d\theta^2+\sin^2\theta d\phi^2$
is the line element on the unit sphere, $0\le \phi < 2\pi $ and
$0\le \theta \le \pi $. Throughout this paper the Newtonian constant $G$ and
the velocity of light $c$ are put equal to 1.

We will study the propagation of polar and axial modes of the quadrupole
gravitational waves (GW thereafter) and the dipole electromagnetic waves
(EW) in the Schwarzschild background. The evolution equation
has the form
\begin{equation}
(-\partial_t^2+\partial_{r^*}^2)\Psi = V\Psi .
\label{2}
\end{equation}
Here $r^*=R+2m\ln \Bigl( {R\over 2m}-1\Bigr) $ is the tortoise coordinate
while the potential term reads: for the polar GW
\begin{equation}
V(R)=6{\eta^2_R\over R^2} +
\eta_R {63m^2(1+{m\over R})\over 2R^4(1+{3m\over 2R})^2};
\label{3}
\end{equation}
for the axial GW
\begin{equation}
V(R)=6{\eta_R\over R^2}(1-{m\over R});
\label{4}
\end{equation}
and for the dipole EW
\begin{equation}
V(R)=2{\eta_R\over R^2}.
\label{5}
\end{equation}
The evolution equations corresponding to the first two potentials
 are called the Zerilli equation \cite{Zerilli}  and the Regge-Wheeler
 equation \cite{Regge}, respectively.

\subsection{Conserved energy}

The equation (\ref{2}) possesses a conserved energy,
\begin{equation}
E(R,t)=\int\limits_R^{\infty }{dr\over \eta_r}\Biggl( (\partial_t\Psi )^2
+ (\partial_{r^*}\Psi )^2 + V\Psi^2\Biggr) ;
\label{6}
\end{equation}
that is, the rate of change of $E$ in a fixed volume equals to the
total flow through the boundary (\cite{kark2001} and \cite{kark2002}).
This agrees (up to a constant factor)
with the energy deduced from the stress-energy tensor for the
EW. Eq. (\ref{6}) represents a mathematically useful quantity
in the case of the gravitational waves, with the density being asymptotically
proportional to the density in the quadrupole formula. In either case,
the energy  conservation becomes important in our forthcoming
construction.

Assume that initial data $\Psi $ and $\partial_t\Psi $ vanish
inside a sphere having a radius $a>2m$.
From the conservation law one easily finds that the amount of the energy
that reaches a distant observer is equal to
\begin{equation}
E_a(\infty )\equiv  E(a,t=0) -\delta E_a,
\label{7}
\end{equation}
where
\begin{equation}
\delta E_a= \int\limits_0^{\infty } dt \Biggl(
(\partial_t\Psi +\partial_{r^*}\Psi )^2+V\Psi^2\Biggr)
\label{8}
\end{equation}
(\cite{kark2001} and \cite{kark2002}).
The integration in (\ref{8}) is done along the outgoing
null cone that starts from $a$ at $t=0$.
In the Minkowski spacetime (put formally $m=0$ in Eq. (\ref{2})) all
of an  initially outgoing radiation would get to infinity; in this
case $\delta E_a=0$, since there is no diffusion through the null cone that expands
outward from the initial position $R=a$. It is meaningful to distinguish
between the momentarily  outgoing and ingoing radiation also in a curved,
 but asymptotically flat, spacetime.
One can give either an operational  or an analytic
definition. Imagine a directional wave
generator that sends all radiation in a fixed direction, when located
in an almost flat region. (That makes sense, since it is known from analytic
estimates, that the fraction of the backscattered energy must fall off at least as
$C\times (2m/a)^2$, where $C$ is of the order of unity -- \cite{kark2001},
\cite{kark2002} and \cite{Gerhard}. By choosing a sufficiently distant location
one can make the diffused energy $\delta E_a$
 arbitrarily small.) This generator, when
carefully moved    to a strongly curved region, will preserve its property
of generating directed radiation, which can be initially purely outgoing
(or initially purely ingoing). Alternatively, one can work out an analytic
definition. Initial data can  always be split  into two parts, one
``initially outgoing'' (defined below; in the Minkowski spacetime that would
all get to the infinity) and the other purely ingoing (its form is similar
to the former -- just change $r^*-t$ into $r^*+t$ and some signs in the
expansion -- but it is purely ingoing in the Minkowski spacetime).
We will show in Sec. 3, that the concept of initially
 outgoing waves   is  useful in bulding a nontrivial construction, and that
fact in itself justifies  this  notion.

\subsection{Initial data for  wave equations}

Let us define
\begin{equation}
\tilde \Psi (R,t)=\Psi_0(r^*-t)+{\Psi_1(r^*-t)\over R}+
{\Psi_2(r^*-t)\over R^2},
\label{9}
\end{equation}
where $\Psi_i$'s ($i=0,1,2$) satisfy following relations:
\begin{eqnarray}
&&\partial_t\Psi_1=3\Psi_0,~~~~~~~~\partial_t\Psi_2=\Psi_1-m\partial_t\Psi_1,
\nonumber \\
&&\partial_t\Psi_1=3\Psi_0,~~~~~~~~\partial_t\Psi_2=\Psi_1-{m\over 2}\partial_t\Psi_1,
\nonumber \\
&&\partial_t\Psi_1=\Psi_0,~~~~~~~~~~~~\Psi_2=0,
\label{10}
\end{eqnarray}
for the polar GW, axial GW and dipole EW, respectively. In the Minkowski spacetime
the function $\tilde \Psi $ exactly solves Eq. (\ref{2}). We assume that
$\tilde \Psi_i(r^*,t=0)$'s vanish for  $R\le a$. Notice that only one
of the three functions (for instance $\Psi_0$) can be freely chosen.

We will say that initial data are purely outgoing if on the initial
hypersurface $\Psi =\tilde \Psi $ and $\partial_t\Psi =
\partial_t\tilde \Psi $.
The full solution of Eq. (\ref{2}) can be now split into the known part
$\tilde  \Psi $ and an unknown $\delta $,
\begin{equation}
\Psi =\tilde \Psi +\delta ,
\label{11}
\end{equation}
with null initial values for $\delta $ and $\partial_t\delta $.
$\delta $ is evolved according to the
inhomogeneous wave equation
\begin{equation}
(-\partial_t^2+\partial_{r^*}^2) \delta = V\delta +\tilde V,
\label{12}
\end{equation}
where

\begin{eqnarray}
&&\tilde V = \Biggl( V-6{\eta^2_R\over R^2}\Biggr) \Biggl(
\Psi_0 +{\Psi_1\over R}+{\Psi_2\over R^2}\Biggr) +\nonumber \\
&&{2m\eta_R\over R^4}
\Biggl(-3\Psi_1+2{\Psi_2\over R}\Biggr) ,
\label{13a}
\end{eqnarray}
for polar gravitational modes,
\begin{eqnarray}
\tilde V =
{10m\eta_R\over R^5}
  \Psi_2 ,
\label{13}
\end{eqnarray}
for axial GW and
\begin{equation}
\tilde V=6m{\eta_R\over R^4} \Psi_1
\label{14}
\end{equation}
for the electromagnetic case. The splitting (\ref{11})
has been crucial in obtaining  the analytic estimates of
the backscatter \cite{kark2001} -- \cite{Malec} but
it appears to be advantageous also from the numerical
point of view.

\section{Extremizing the diffusion parameter}

\subsection{Diffusion parameter and the variational problem}

Let us define the {\bf reprocessed radiation } (RR)
as  that   reaching a distant observer after the passage of the
initial pulse; the delay is caused by  multiple backscatterings.
RR would be absent in the Minkowski spacetime.
(For an example, see  on Figs 8-13  the parts of waveforms
to the right from $x=0$.)
We study hereafter the {\bf RR} generated
by initially outgoing waves, in order  to separate the genuine
effects of the geometric curvature  from those
implied by artificial initial data.

The diffusion parameter $\kappa $ is defined as  the ratio of the diffused
energy and the initial energy,
\begin{equation}
\kappa = {\delta E_a\over E(a,0)}.
\label{15}
\end{equation}
Our aim in this section is to provide outgoing initial data that
maximize $\kappa $. This will be done in a class of data that {\bf
do vanish for $R\le a$}.   The intuition behind this is that
if $\kappa $ is large then the fraction of the energy
of the {\bf reprocessed radiation }  should be also large.  That in turn should
translate into effects like vigorous  ringing modes or tail terms.
We conjecture, that there exists a correlation between
$\kappa $ and  (defined in some way) the strength of   QNM's.

Expressing things in technical terms: we want to maximize the
nonnegative quadratic form $\delta E_a$ while keeping
fixed the positive quadratic form $E(a,0)$. In numerical
calculations this task reduces to a multi-dimensional
algebraic eigenvalue problem, as we shall demonstrate.
In the first step we choose some large $R_1\gg a$ -- the
upper end of the initial support -- and maximize $\kappa $
in the future  domain of dependence of $(a,R_1)$ with
the appex at   $(R_2\equiv R({r^*(R_1)+r^*(a)\over 2}),
t={r^*(R_1)-r^*(a)\over 2})$. Obviously the
change of $R_1$ would change $\kappa $ as well, but
it has been established that above some critical value of
$R_1$ the value of  $\kappa $ stabilizes.
It has been found by the method of trial and error
that the choice $R_1\approx 150m$ is  satisfactory.

\subsection{Discretization of the variational problem}

In the second step we determine a functional discrete basis
$\{ f_i\} $ ($i=1...N$) on the closed interval $[r^*(a), r^*(R_1)]$.
The dimension of that basis was usually 250
(but tests with smaller and bigger dimensions were
also  done) -- a number much smaller than the number
of points (8000) in the spatial grid; that
facilitated greatly the numerical calculation,
without loosing  accuracy.  The best results have
been obtained for the basis consisting of the first
250  Legendre polynomials with odd indices.

Let the expansion of the function $\Psi_0$ (the only free function
in the initial data set -- see the remark folowing Eq.
(\ref{10})) be
\begin{equation}
\Psi_0(r^*,t)=\sum_{i=1}^NC_if_i\Bigl( {r^*-t-r^*(a)\over r^*(R_1)-r^*(a)}
\Bigr) .
\label{16}
\end{equation}
Then one finds from Eq.   (\ref{6}) that the total initial
energy is a positive definite quadratic form,
\begin{eqnarray}
&&E(a, R_1, t=0)=\nonumber \\
&&\int\limits_a^{R_1}{dr\over \eta_r}\Biggl( (\partial_t\Psi )^2
+ (\partial_{r^*}\Psi )^2 + V\Psi^2\Biggr) |_{t=0} = \nonumber \\
&&\sum_{i,j=1}^N B_{ij}(a, R_1) C_iC_j,
\label{17}
\end{eqnarray}
where the matrix $B_{ij}$ is known from the  numerical calculation.

Each element $\Psi_{0,i}= f_i$ determines some initial
values $(\Psi_{f_i})_{t=0},(\partial_t\Psi_{f_i})_{t=0}$;
they give rise to solutions $\Psi_{f_i}$ in the domain
of dependence. These solutions are linearly independent
(due to the uniqueness of solutions of Eq. (\ref{2})).
Therefore the solution generated by the initial data defined
by (\ref{16})  can be expressed as   the
linear combination
\begin{equation}
\Psi (r^*,t)=\sum_{i=1}^NC_i\Psi_{f_i}.
\label{18}
\end{equation}
Thus the  energy $\delta E_a$ diffused through the null cone
connecting $(a,0)$ with $(R_2,t)$ has the form
\begin{equation}
\delta E(a, R_2)=\sum_{i,j=1}^N A_{ij}(a, R_1) C_iC_j.
\label{19}
\end{equation}
Again the matrix $A$ is obtained numerically.

The task of maximizing the ratio of the two quadratic forms is
equivalent to finding eigenvalues in the generalized eigenvalue
problem
\begin{equation}
\sum_{j=1}^NA_{ij}C_j=\lambda \sum_{j=1}^NB_{ij}C_j,
\label{20}
\end{equation}
where $\lambda $ is the eigenvalue and $(C_i)$ is
the corresponding eigenvector.
There are many excellent numerical procedures for solving
of the generalized problem. We choose one from the fast
EISPACK package. This allowed us to find several
largest eigenvalues $\lambda ^{(k)}$, the
eigenvectors $(C_j^{(k)})$ and, from Eq. (\ref{16}), the
corresponding functions $\Psi_0^{(k)}(a, R_1,N)$ for
 $k=1,2...$. Having $\Psi_0^{(k)}$ one finds initial
 data using Eqs (\ref{9}) and  (\ref{10}).

As a consistency check, in number of cases
the wave packet given by
 $\Psi_0(a, R_1,N)$ was evolved and the diffusion parameter
was found directly from the definition. In the case
of disagreement the procedure
could  be repeated with other values of numerical parameters.
The disagreement was never observed for the vectors  maximizing
$\kappa $, but it was found in number of cases with
fourth and fifth eigenvectors (by convention, the  eigenvectors
are ordered according  to the decreasing eigenvalue,
$\lambda_1>\lambda_2>...$).
The parameters ($N=250,~ r^*(R_1)\approx r^*(a)+160m$,
the size of the grid) that
are reported above seem to be optimal, in the sense that the
corresponding integration
time was not too long  while   the accuracy was kept
reasonably good. These values  have been obtained by
performing many series of numerical calculations.

\subsection{Final preparation of extremal initial data}

These pre-prepared initial data that {\bf are maximizing
within the chosen region} (in the future dependence zone of data
defined on ($a, R_1)$) undergo a process of extending
the initial data beyond $R_1$. Strictly saying, we match
a function
\begin{equation}
f(r^*)=C_2+C_1\exp (-r^*/10)
\label{21}
\end{equation}
to each eigenvector  $\Psi_0^{(k)}(a, R_1,N)$.
The matching is differentiable and the gluing point $G$
is selected  independently for each eigenvector.
The value of $G$ has been obtained as follows.  Fixing $a$ and
$R_1$, one finds  $\Psi_0^{R_1}$ (the upper index is put here in order
to stress the local character of the procedure)   and
initial values of the   locally extremizing solution $\Psi^{R_1}$.
With the increase of $R_1$, while  keeping $a$ fixed, the function
$\Psi_0^{R_1}$ changes. In the limit  one should  in principle
obtain the sought
extremizing solution,  $\Psi =\lim_{R_1\rightarrow \infty } \Psi^{R_1}$.
In the numerical  practice the integration region must be finite.
The dependence of $\Psi_0$ on $R_1$
suggests that  $   \Psi_0^{R_1}\approx const$
outside some region of compact support. The point $G$ is numerically
determined as being  some point near the transition region.
In our cases we obtained  $r^*(8m)\le r^*(G)\le  r^*(25m)$.
Therefore the chosen $\Psi_0$   bears on an asymptotically constant value.
   Fig. 1 shows   initial profiles of  the the third eigenvector
$\Psi_0$  for $a=2.001$ (GW, the polar mode).

\begin{figure}[1]
\epsfxsize=6cm
\centerline{\epsffile{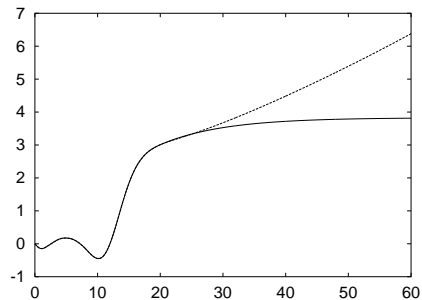}}
\caption{ An eigenvector is matched to an asymptotically constant
function (lower branch -- solid line). The upper branch (broken line)
represents the eigenvector before matching.
 The x-axis shows $r^*-r^*(a)$ and
is scaled in units of $m$.}
\end{figure}

We would like to point that this process of  matching
is to  a degree arbitrary and   the obtained eigenvectors
can be expected to be close (but not necessarily identical)
to the extremizing eigenvectors.

\section{Numerical results}

\subsection{Extremizing initial data and  $ \kappa $ versus
 $a$}

 Figs. 2 and 3 show the distribution of the initial
energy densities of the first and the fifth  axial GW  modes.
 As one can expect,
the mass center is closer to the horizon in the case of  the
extremal data, while (not so obviously) the graph of the fifth
vector suggests a larger contribution of high frequency radiation.
\begin{figure}[3a]
\epsfxsize=6cm
\centerline{\epsffile{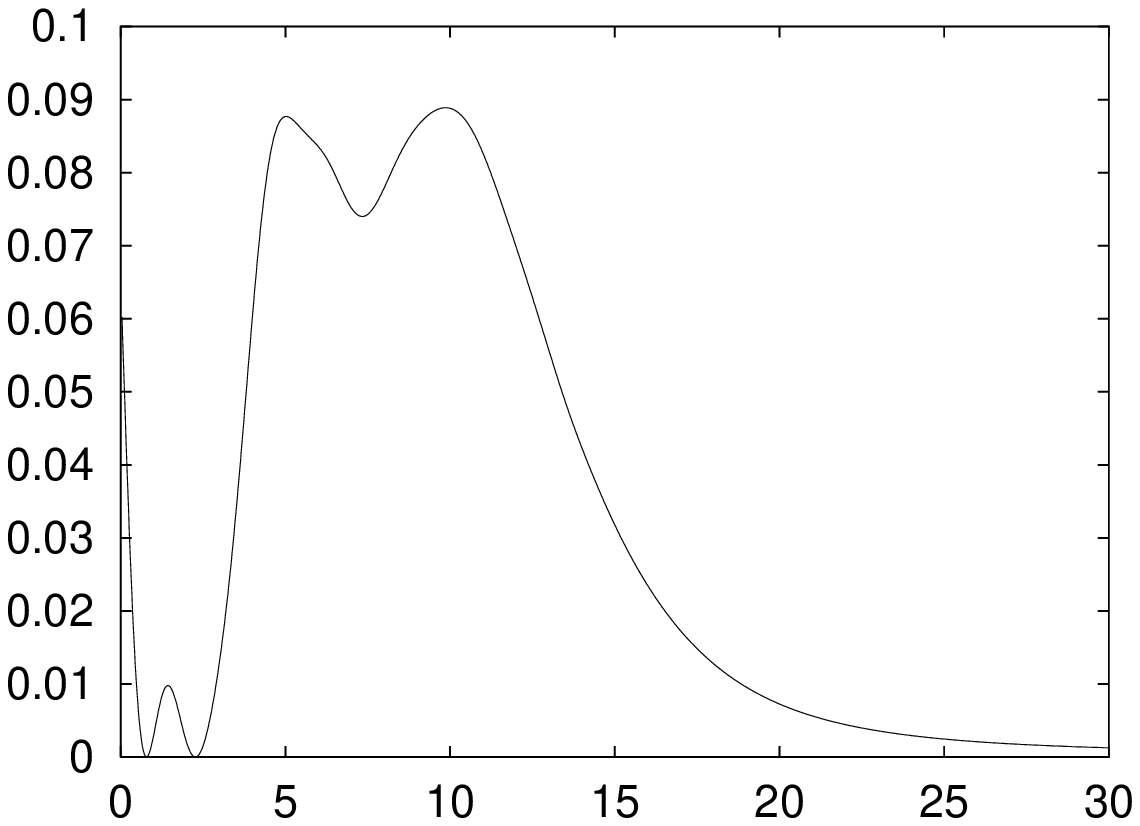}}
\caption{  }
\end{figure}
\begin{figure}[3b]
\epsfxsize=6cm
\centerline{\epsffile{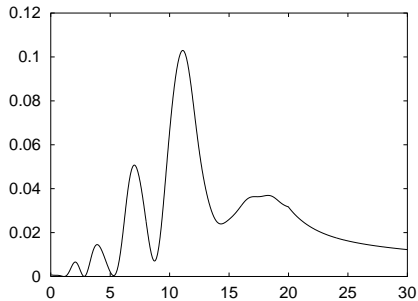}}
\caption{ Axial GW.  Initial energy densities  for the 1st (Fig. 2) and 5th
(Fig. 3) eigenvectors. Here $a=2.01$.  The x-axis shows values
of $r^*-r^*(a)$ and
is scaled in units of $m$. }
\end{figure}
Fig. 4 demonstrates that the energy support of maximizing initial
data increases with  the increase of $a$. The larger
$a$ the larger  distance at which the value of the
energy stabilizes. That feature of the maximal initial data is
counter-intuitive at the first glance, since the backscatter
is strongest  in regions with large values of the potential $V$
(around $R\approx 3m$ or $r^*\approx 0$) and one would
 expect accumulation of the energy near $a$ if $a\gg 2m$. The reason why it
is not so is that the backscatter depends also on the frequency;
a radiation accumulated at $a$ would be dominated by  high--frequency
waves, which are weakly backscattered.

\begin{figure}[7]
\epsfxsize=6cm
\centerline{\epsffile{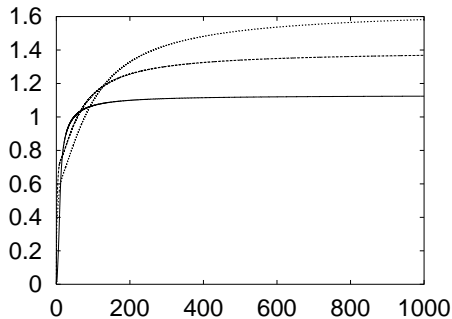}}
\caption{ Polar GW.  Initial energy
$E(R)\equiv E(a,0)-E(R,0)$ ($y-$ axis) as a function of $r^*-r^*(a)$ for
$a=2.1m$ (solid line), $a=3.1m$ (broken line) and $a=4m$ (dotted line).}
The scale of the ordinate is arbitrary while the abscissa is in units
of $m$.
\end{figure}

The main lesson that can be drawn from the foregoing
discussion  is that  the extremizing initial data
can occupy a large region that extends far away from the
black  hole horizon.

A question arises, whether one can have modes with large
$\kappa $ in the case of waves that are initially well separated
from the horizon, i.e., when $a \gg 2m$. As it happens,
in order to give an answer one has to  combine the numerical
approach and an analytic insight.
This is because the numerical time is proportional
(with some large coefficient)
to $(a/2m)^2$ and the numerics is feasible only when $a$
is not too big.  Fortunately, analytic
estimates  show that the diffusion parameter quickly decreases with
the distance, at least as quickly as $(2m/a)^2$, and becomes small
at large $a$ \cite{remark}. Therefore if $a\gg 2m$ then no modes
with large $\kappa $ can exist and it suffices  to restrict
the present analysis to $a$ being relatively small. In this paper
the numerics is done for    $a\le 6m$.

Figs 5 -- 7 show the dependence of $\kappa $ on $a$ for five
succesive eigenvectors
with largest eigenvalues, in each of the considered wave sectors.
While at $a$ very close to $2m$ the largest eigenvalue is close to 1 in all
three cases, then the  eigenvalue fifth in the
order is smaller than 0.01 for
EW and close to 0.1 for polar GW, with the axial GW lying in between.
The next observation that should be made is that with the increase of
$a$, the largest eigenvalue changes slower than the remaining ones
and the the falloff of eigenvalues is quickest for EW and slowest
for the polar GW.
\begin{figure}[2a]
\epsfxsize=6cm
\centerline{\epsffile{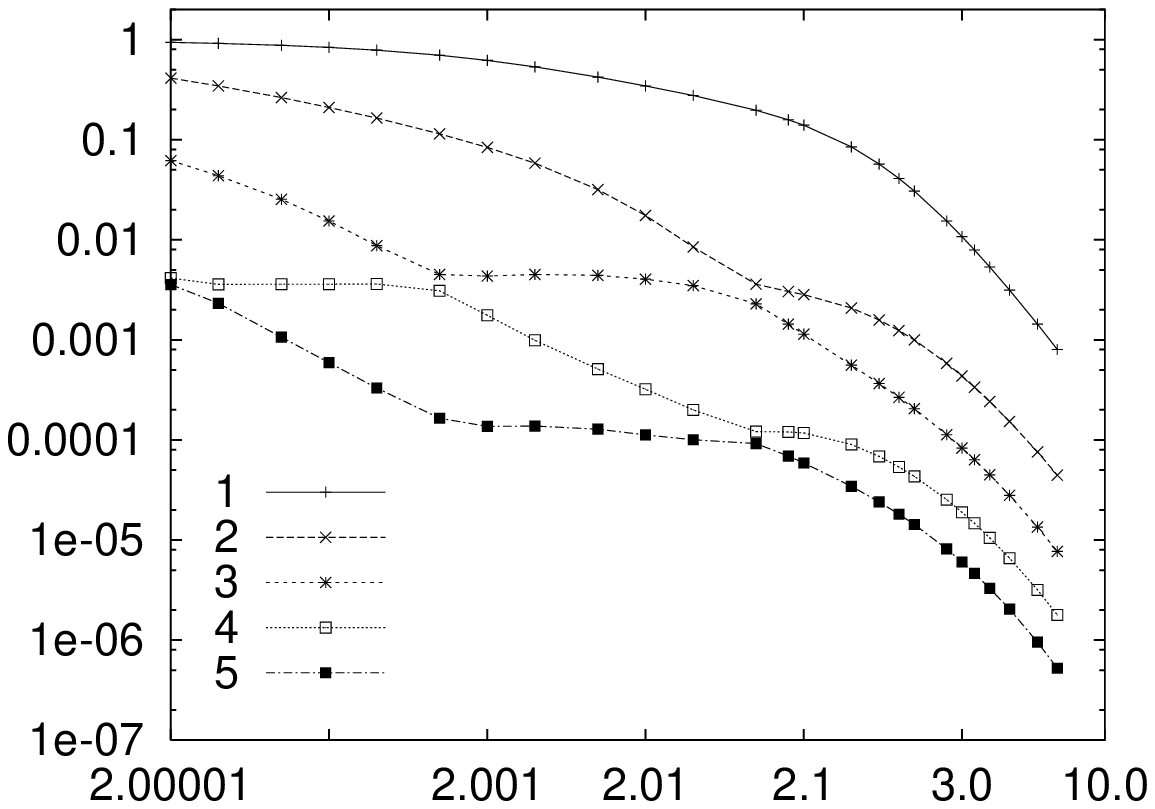}}
\caption{  }
\end{figure}
\begin{figure}[2b]
\epsfxsize=6cm
\centerline{\epsffile{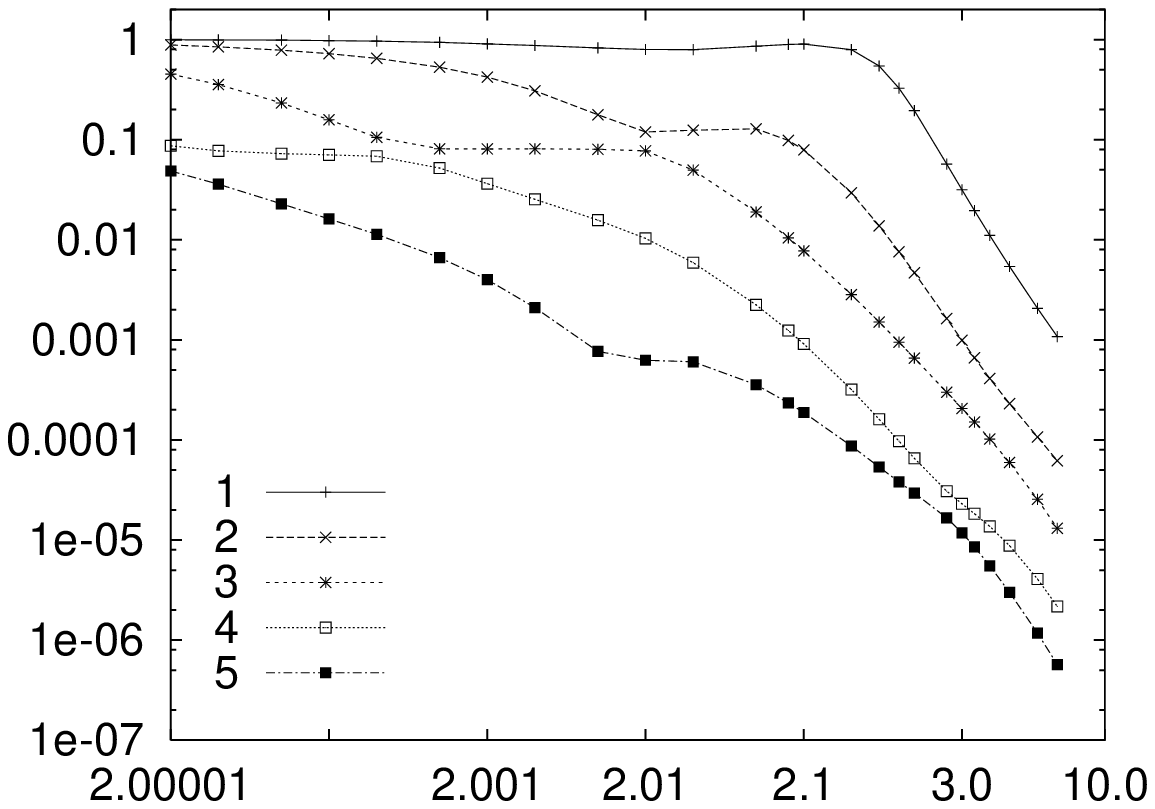}}
\caption{ }
\end{figure}
\begin{figure}[2c]
\epsfxsize=6cm
\centerline{\epsffile{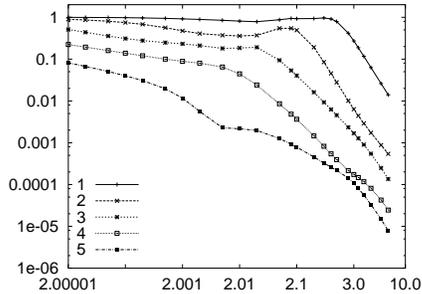}}
\caption{Eigenvalues $\kappa $ for the  first five eigenvectors,
dipole EW (Fig. 5), axial GW (Fig. 6) and  polar GW (Fig. 7),
 in dependence
on $a$. The points are connected by lines (solid, broken, etc.) in order
to make easier the identification of eigenvectors.
  The x-axis shows the position of $a$ and
is scaled in units of $m$. }
\end{figure}

\subsection{On the stability of templates}

Our earlier observation  that  QNM  can be born and can die
\cite{Kark2002}, when  observation points are moved away form the
black hole horizon, can be rephrased as demonstration that
{\it  templates can critically depend
on the distance of an observer  from the horizon}.
 Below we repeat that study and establish
a lower bound on the distance of the observer from the horizon
that is needed in order to detect a reliable wave profile.

 Fig. 8 shows that there are many oscillations
at $R=10m$, which  gradually die when
the observation point is moved away to
$R=100m$ (Fig. 10).  One  can see that only the first eigenvector
produces some distorted oscillations at $R=100m$  while the
remaining two  fail completely to show any ringing.
\begin{figure}[8]
\epsfxsize=6cm
\centerline{\epsffile{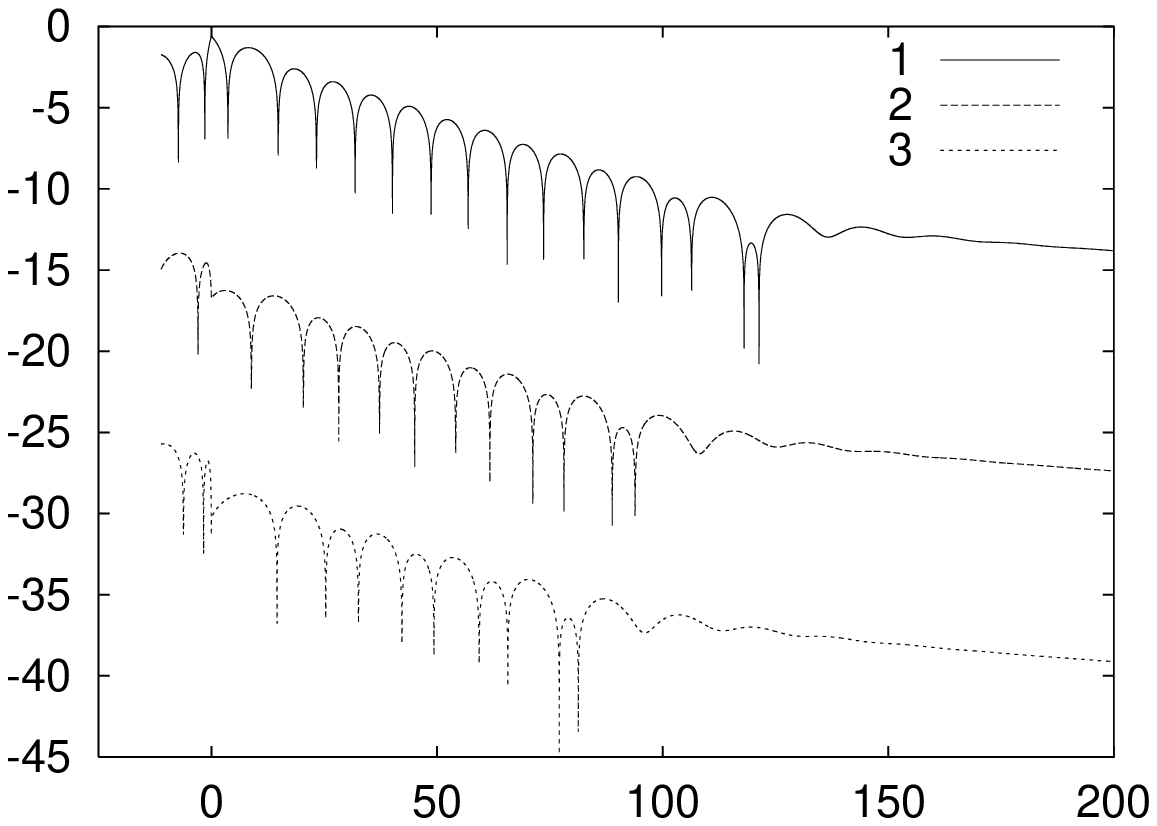}}
\caption{  }
\end{figure}
\begin{figure}[9]
\epsfxsize=6cm
\centerline{\epsffile{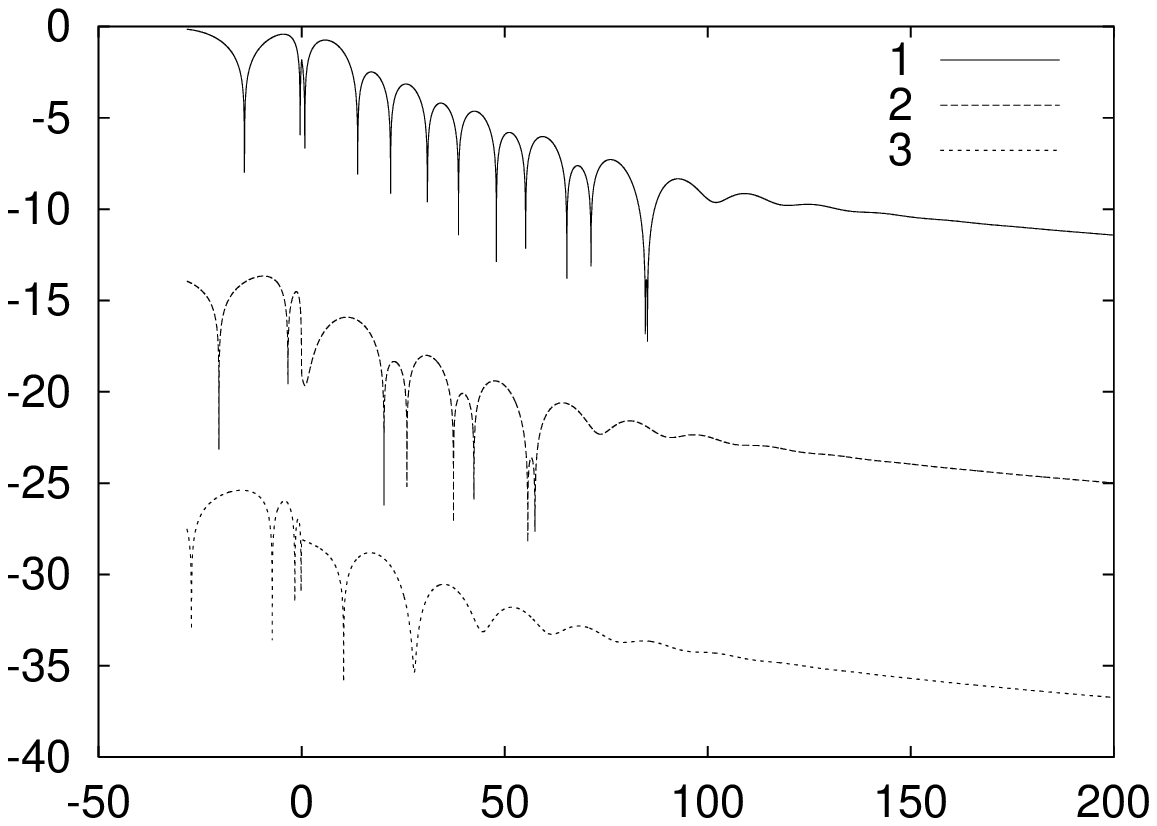}}
\caption{  }
\end{figure}

\begin{figure}[10]
\epsfxsize=6cm
\centerline{\epsffile{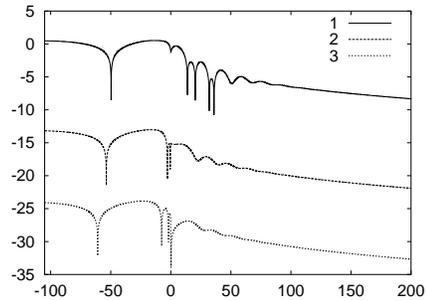}}
\caption{ Polar GW, $a=3m$.  Waveforms ($\ln |\Psi |$)
generated by the three strongest eigenvectors
(no 1 - solid line, no 2 -  broken line and no 3 - dotted line),
taken at $R=10m$ (Fig. 8), $R=25m$  (Fig. 9) and $R=100m$ (Fig. 10).
The abscissa is in units of $m$. The scale of the ordinate is arbitrary
and   the  amplitudes of each type of  eigenvectors are properly
normalized, for the sake of clarity.
The data to the right of    $x=0$ correspond to the purely
backscattered radiation. }
\end{figure}

Notice that while  the  amplitude of the surviving QNM  seems to
increase moderately, the  tail (and pre-tail) part extends and gains
in power significantly. This agrees with conclusions of  \cite{Kark2002}.
Particularly interesting is the comparison of templates shown in Fig. 9,
 taken at $R=25m$   \cite{25m} and in Fig. 10, determined at
$R=100m$. They are clearly different -- the ringing phase can be much
shorter or even disappear, while  the remnants of the initial data
(the parts of the diagrams to the   left from $x=0$) seen at $R=25m$
are completely different from those detected at $R=100m$.
One can conclude  that  {\it the process of taking
waveforms is unstable under the  translation
of the observation point} --   the templates can strongly depend on the
location of the observer.

One can also infer from the preceding information
that $25m$ is too close for being the observation point
and $100m$ may well be the lower bound for the observer's position.
To this point, let us add  that in many of
analyzed examples (not reported here) the waveforms did
 not change significantly above $R=100m$.

\subsection{Strong   ringing modes}

One of  our  aims is to find initial
data that give the strongest possible ringing within the
{\bf reprocessed radiation}.
The  diffusion energy $\delta E_a$ bounds   the energies
of QNM, the tail (and pre-tail)  term and also of the radiation
falling to a black hole.
While we do not have analytic estimates of the shares of the
particular contributing terms in $\delta E_a$, it is obvious that
configurations with large $\kappa $ have some room for robust
oscillations.  For that reason  we study waves defined
by the  {\bf extremal initial data}.

Figs 11 -- 13 present the radiation corresponding to the
EW and GW initial pulses as seen by an  "observer"
situated at $R=100m$.   The $x=0$ point of
the abscissa corresponds to the  moment of time
$t=r^*(100m)-r^*(a)$. This train of  data that moves with the speed of
light is seen earlier ($t<100m-r^*(a)$)  and it lies to the left
from $x=0$. To the right  from $x=0$ we have $t>100m-r^*(a)$; in the
absence of  the backscattering there would be no signal
at all.
\begin{figure}[11]
\epsfxsize=6cm
\centerline{\epsffile{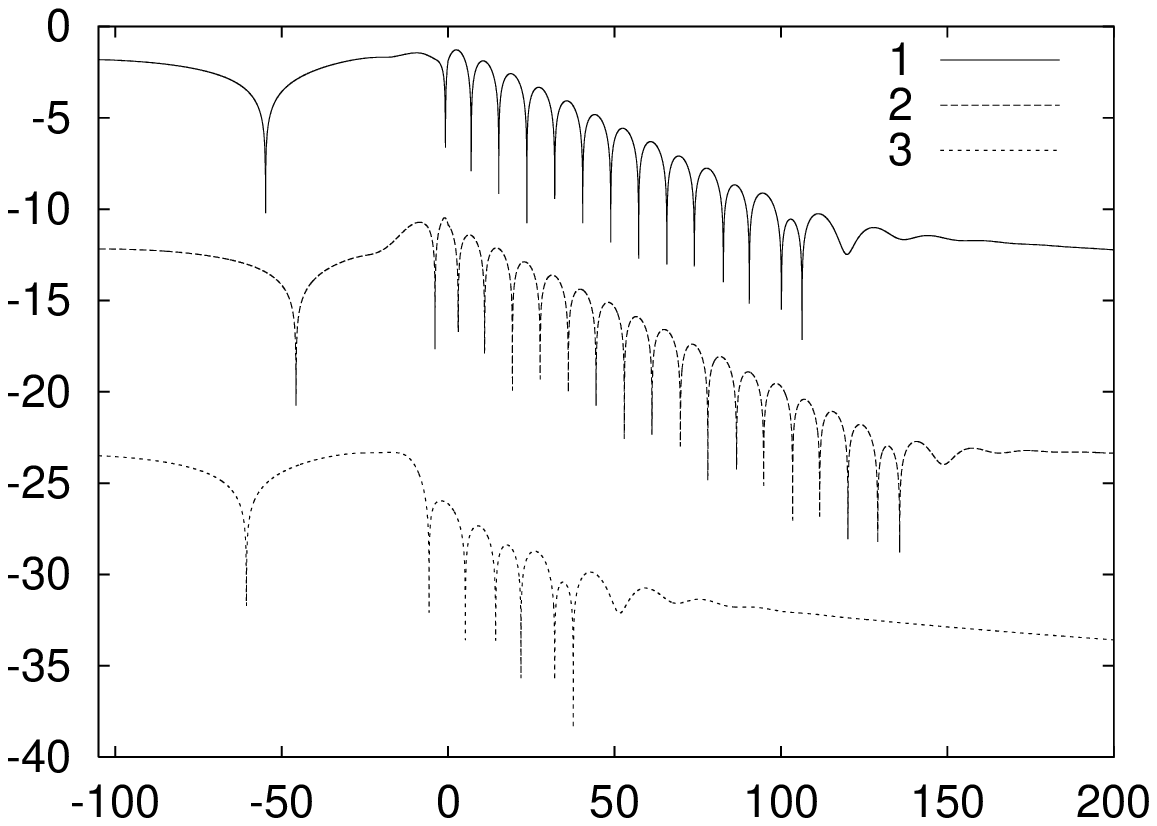}}
\caption{  }
\end{figure}
\begin{figure}[12]
\epsfxsize=6cm
\centerline{\epsffile{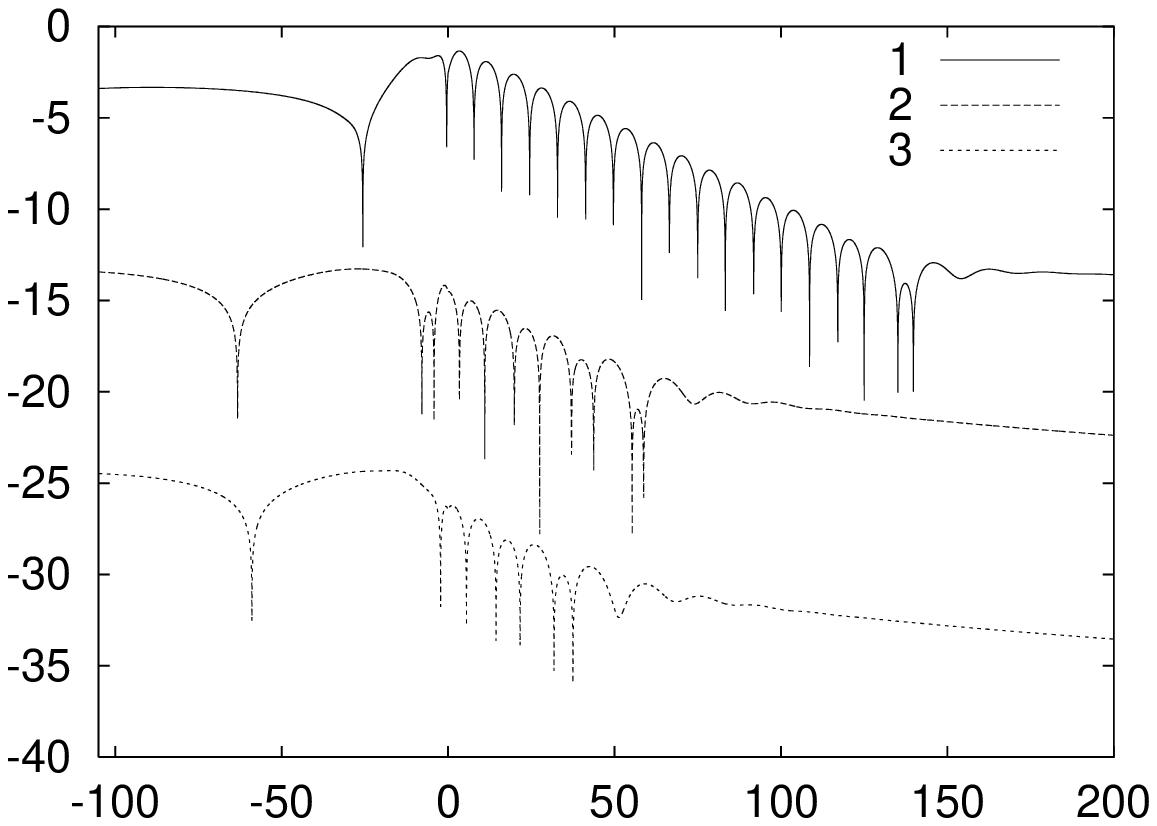}}
\caption{  }
\end{figure}
\begin{figure}[13]
\epsfxsize=6cm
\centerline{\epsffile{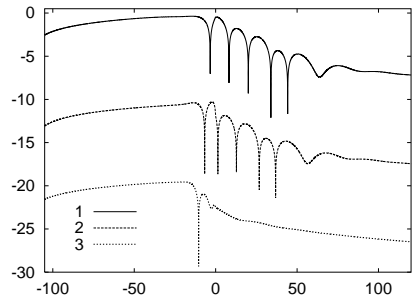}}
\caption{Templates ($\ln |\Psi |$) of the axial GW (Fig. 11), polar GW (Fig. 12)
and EM waves (Fig. 13), the three strongest eigenvectors (depicted  as in
Figs 8 -- 10). The "observer" is located at $R=100m$ and the parameter
$a=2.001$. The part to the right of    $x=0$ is the purely
backscattered radiation. }
\end{figure}
 Notice that
 the amplitudes of the strongest ringing mode
 are of the order of  the largest   amplitudes of remnants of
the original radiation. This is particularly
clearly manifested  in the case of the strongest
polar  GW eigenvector.  Observe also a strong deformation of the
original signal just before $t=r^*(100m)$; initial waveform
would be zero at $x=0$, while in  Figs 10 -- 13 one can see a gradual
build-up of a backscattered signal. Again the effect is strongest
for the   polar GW (Fig. 12, the 1st eigenvector), when the
 backscattered part exceeds the remaining signal by a factor of 10.
We would also like  to direct the attention of the reader to
Fig. 10. There the ringing is absent for the second and third
eigenvector, but  a very strong pre-tail
term is observed, comparable to the remainder of the
main signal.

These examples essentially  confirm our conjecture
that there exists a correlation
between the diffusion factor and some features (strongest
QNM and/or the longevity of the ringing phase) of the ringing.
(The reservation "essentially" is caused by the fact, that the ringing
belonging  to the second axial mode on Fig. 12
is stronger than that of the  first eigenvector; but in this case the
diffusion parameters differ only by the factor of 2.)
An intuitive  explanation with  analytic flavour  would be following.
There is effectively $(\partial_t\Psi )^2$ contribution
to the observed energy flux, if an observation point is located far
away from the horizon (the asymptotic zone, where
  radiation is  dominated by the
$\Psi_0$-type term and $\partial_t\Psi \approx -\partial_{r^*}\Psi $).
 Quasinormal modes oscillate and therefore
they give a more significant contribution to the total
backscattered  energy  than, say, tail terms. Hence small $\kappa $
would be prohibitive for any ringing, while strong $\kappa $
leaves this possibility open. This reasoning suggests also that
the diffused energy might well be the best measure  (imperfect,
admittedly) of the energy of quasinormal modes generated by moving
wave pulses.

It was reported earlier (see, for instance, Sec. IX
in \cite{Andersson1994}) that there exists a (sharp value of)
critical width (suitably defined) of initial data corresponding to strong ringing
and that both (sub- and super-) critical data generate much weaker
ringing. While we observe a kind of a similar dependence, it is certainly
less dramatic and no sharp indicator seems to be appropriate.
Admittedly, we deal with a different situation -- there is
only an (initially)  outgoing radiation, while
in \cite{Andersson1994} there are both (initially)
ingoing and outgoing components -- but that probably is not
relevant. More important can be a different shape of initial data
-- here determined by the extremization procedure of the preceding section,
while in  \cite{Andersson1994}  assumed to be gaussian.

\section{Linear versus nonlinear   descriptions of the post-merger
evolution}

\subsection{How typical are ringing modes }

That is a basic tenet of the General Relativity, dictated by
the beliefs in the cosmic censorship \cite{penrose} and no-hair
conjectures \cite{mazur},
that at  some stage after plunge/merger a geometry generated by a
pair of black holes can be represented as a single perturbed
 black hole. The perturbations would be represented by
gravitational waves and the final black hole would be either
spinning (the Kerr  black hole) or nonspinning (the  Schwarzschild
black hole), the latter in the case of the  head-on collision.
The so-called close limit approximation (\cite{Pullin99},
\cite{Anninos}, \cite{Gleiser}) seems to assert that the linear approximation
is valid since the formation of  a common  apparent horizon.
Anninos et al.\cite{Anninos} give  some arguments in favour of
this claim  that are supported (albeit with some reservations)
by their analysis of the head-on collisions  \cite{Smarr},
with initial data of Misner type  \cite{Misner}. 
Gomez et al. \cite{Gomez}  provide  other supporting arguments
in their discussion of fissioning white holes. An interesting new 
feature of a recent work by Husa et al. \cite{Husa}, which uses close
approximation, is a weak dependence of waveforms on the 
collision velocity of two black holes.

If this scenario is right  then  the naive expectation would be that
most of the radiation is concentrated in the  vicinity of the  horizon
\cite{remarkenergy}. (This is in fact observed -- see Fig. 10 in
\cite{Anninos}, which shows that initial perturbation extends to regions
very close to $R=2m$.) One can split these initial data into initially
ingoing and outgoing parts, according to the descriptions of Secs 2.B
and 2.C. (In the example given in Fig. 10 of \cite{Anninos}
the ingoing radiation  remains forever inside the potential well
 \cite{remark1}.)  The latter
can be expanded in the  diagonalizing basis  defined in Sec. III
and consisting of 250 base vectors, with
the parameter $a$ (that in fact specifies  this basis - see
Sec. III)  being very close to $2m$. But if $a\approx 2m$, then
Sec. IV.A suggests that there appear a number of
eigenvectors (from at least 2 for EW to at least 4-5 in the case of polar GW)
with  diffusion parameters being close to 1.
The initial data  for the linear phase are determined by the
preceding  nonlinear evolution; if these were purely  random,
then the chance  of having large $\kappa $ (and
strong ringing) would be of the order of 1\%.
Leaving aside the question whether the merger phase can be
regarded as a random process, the least one can say is that the  maximizing
initial data of Sec. III should not be  apriori excluded.
Another argument is that the process of the backscatter is selective
 --  waves longer than QNM's are
stronger  backscattered. Since the present gravitational wave detectors
are tuned to frequences smaller (even  by one order in the case of less
massive binaries of black holes \cite{Pullin99})
than QNM's characteristic for the most likely sources of the gravitational
radiation, there are reasons to expect that the detected radiation will be
strongly backscattered (even if the ringing itself would be undetectable).

\subsection{Taking of waveforms -- some parameters}

We pointed out  in Sec. IV. B that in some examples
the templates are unstable with  respect to the change of the ``observation''
point. No significant changes in the wave profiles
have been observed beyond $100m$, but the wave profiles at $R=25m$
can be profoundly different from those taken at the former point.
 This suggests that the determination of
templates should    be done at $R\gg 25m$ and that
 $R=100m$ might the smallest required distance of the observer.
In some cases    the  duration of the remnants of the
 original signal and of the comparable in strength
 backscattered part
exceeds   $150m$ (see Fig. 12).
From this one can infer that the minimal integration time must
exceed at least $150m$; it happens to be the
 longer the closer the initial pulse is
 located to the horizon of a black hole.
There is no numerical calculation,  in the case of  the full
nonlinear  collisions, up to our knowledge, which satisfies both 
requirements. For instance in \cite{Anninos}, the observations were held at
$R=25m$ through the time interval $100m$.

\section{Concluding remarks}

The Schwarzschild spacetime can be believed to provide a good approximation
to the last phase of the collapse of two black holes -- the so-called close
limit (\cite{Pullin99} -- \cite{Gleiser}) bases on this idea --
provided that the collision is (almost) head-on. At this stage the spacetime
can be regarded as consisting of a single black hole having a mass $m$ and
some  gravitational  radiation that propagates  on a Schwarzschlidean
background. The  initial data for the  linear evolution of the
gravitational radiation  should be provided by a numerical
solution of the preceding phases of the collapse. This task is at
present (and presumably for some years to come) unavailable for numerical
relativists. In this   context  the existence of universal
{\bf imprints} of the spacetime curvature like {\bf QNM's} could be of
relevance, but only if their amplitudes are strong enough.

We invent  a variational procedure that generates initial data corresponding
to the strong backscatter. These initial data have some features that might
 look as counter-intuitive; they have an extended support and a significant
fraction of the wave signal energy comes from a distant region ($R\gg 2m$).
 They   generate strong ringing modes or (if the ringing is absent)
robust  terms preceding the tail. In many cases  the backscattered terms
and QNM's are  much stronger than the remnants  of the original signal;
therefore the  QNM's waveforms cannot be ruled out as objects of interest
for the gravitational wave astrophysics.

Finally, it has been shown  how from the linear description
one can get clues as to the preparation of templates in
the numerical analysis of the full nonliner problem
of collapsing (head-on) black holes.

Acknowledgments. EM thanks Bernd Schmidt,
Edward Seidel and Jonathan Thornburg for discussions
and interesting remarks.  This work has been suported
in part  by the KBN grant 2 PO3B  006 23. ZS thanks the
Pedagogical University for the research grant.


\begin{references}
\bibitem{25m} $R=25m$ seems to be the canonical choice for the determination
of   waveforms   - see \cite{Anninos}.
\bibitem{remarkenergy} This is plausible and implicitly always assumed
to be true by numerical relativists but remains  unproven.
The main difficulty has its roots in the absence
of the notion of the local energy density for free  gravitational fields.
\bibitem{remark1} The r. h. s. of  Eq. (\ref{2}) becomes very small
 when $r^* \ll -1$ and the ingoing part is not backscattered
 near the horizon.
\end{references}
\end{document}